\documentclass[aps,pra,nofootinbib,twocolumn,floatfix,showpacs,superscriptaddress]{revtex4-1}
\usepackage[utf8]{inputenc}
\usepackage{amsmath,graphicx,epsfig,amsthm,color,bm}
\usepackage{pifont,bbding,amssymb,soul}
\usepackage{indentfirst}
\usepackage{times}
\usepackage{appendix, comment}

\usepackage{CJKutf8}
\usepackage{ucs}
\usepackage[encapsulated]{CJK}

\usepackage{tikz,pgfplots}

\newcommand{\braket}[1]{\langle #1 \rangle}

\newcommand{\tr}{{\rm tr}}

\newcommand{\Kcal}{\mathcal{K}}

\newcommand{\abs}[1]{\lvert #1\rvert}

\newcommand{\bra}[1]{\langle #1|}
\newcommand{\ket}[1]{|#1\rangle}
\newcommand{\bracket}[2]{\langle #1|#2\rangle}

\usepackage[colorlinks=true,citecolor=blue,urlcolor=blue]{hyperref}

\usepackage[capitalize]{cleveref}

\begin{document}

\title{Experimental investigation of quantum uncertainty relations with classical shadows}

\author{Lu Liu}
\affiliation{School of Physics, Shandong University, Jinan 250100, China}

\author{Ting Zhang}
\affiliation{School of Physics, Shandong University, Jinan 250100, China}

\author{Xiao Yuan}
\email{xiaoyuan@pku.edu.cn}
\affiliation{Center on Frontiers of Computing Studies, Peking University, Beijing 100871, China}

\author{He Lu}
\email{luhe@sdu.edu.cn}
\affiliation{School of Physics, Shandong University, Jinan 250100, China}

\begin{abstract}
The quantum component in uncertainty relation can be naturally characterized by the quantum coherence of a quantum state, which is of paramount importance in quantum information science. Here, we experimentally investigate quantum uncertainty relations construed with relative entropy of coherence, $l_1$ norm of coherence and coherence of formation. In stead of quantum state tomographic technology, we employ the classical shadow algorithm for the detection of lower bounds in quantum uncertainty relations. With an all optical setup, we prepare a family of quantum states whose purity can be fully controlled. We experimentally explore the tightness of various lower bounds in different reference bases on the prepared states. Our results indicate the tightness of quantum coherence lower bounds dependents on the reference bases as well as the purity of quantum state.  
\end{abstract}

\maketitle

\section{Introduction}
The uncertainty principle lies at the heart of quantum mechanics that differs it from classical theories of the physical world. It behaves as a fundamental limitation describing the precise outcomes of incompatible observables, and plays a significant role in quantum information science, from quantum key distribution~\cite{Koashi2006,Koashi2009,Berta2010,Tomamichel2011} to quantum random number generation~\cite{Vallone2014,Cao2016}, and from quantum entanglement witness~\cite{Prevedel2011,Li2011,Berta2014}
to quantum steering~\cite{Walborn2011,Schneeloch2013} and quantum metrology~\cite{Giovannetti2011,Hall2012} (also see Ref.~\cite{Coles2017} for the review of uncertainty relation and applications).

The seminal concept of uncertainty relation was proposed by Heisenberg in 1927~\cite{Heisenberg1927}, in which he observed that the measurement of position $x$ of an electron with error $\Delta(x)$ causes the disturbance $\Delta(P)$ on its momentum $p$. In particular, their product has a lower bound set by Planck constant, i.e., $\Delta(x)\Delta(p)\sim\hbar$. Later, Robertson generalized the Heisenberg's uncertainty relation to two arbitrary observables by $\Delta A\Delta B\geq\frac{1}{2}|\langle [A,B]\rangle|$ with $\Delta A$ ($\Delta B$) being the standard deviation of observable $A$ ($B$), $[A,B]=AB-BA$ being the commutator of $A$ and $B$ and $\braket{\cdot}$ being the expected value in a given state $\rho$~\cite{Robertson1929}. Indeed, such a uncertainty relation has a state-dependent lower bound so that it fails to reveal the intrinsic incompatibility when $A$ and $B$ are non-commuting.     

To address the issue of state-independence of Robertson's uncertainty relation, the entropic uncertainty relation has been developed by Deutsch~\cite{Deutsch1983}, Kraus~\cite{Kraus1987} and Maassen and Uiffink~\cite{Maassen1988}: Consider a quantum state $\rho$ and two observables $A$ and $B$, the eigenstates $\ket{a_i}$ and $\ket{b_i}$ of observable $A$ and $B$ constitute measurement bases $\mathbb{A}=\{\ket{a_i}\}$ and $\mathbb{B}=\{\ket{b_i}\}$. The probability of measuring $A$ on state $\rho$ with $i$th outcome is $p_i=\tr[\rho\ket{a_i}\bra{a_i}]$, and the corresponding Shannon entropy of measurement outcomes is $H(A)=-\sum_ip_i\log_2p_i$. Then, $H(A)+H(B)$ is lower bounded by $H(A)+H(B)\geq -\log_2 c$ with $c=\max_{i,j}|\bracket{a_i}{b_j}|$ is the maximal overlap between $\ket{a_i}$ and $\ket{b_j}$. According to the definition of Shannon entropy, $H(A)$ quantifies the uncertainty or lack of information associated to a random variable, but does not indicate weather the uncertainty comes from classical or quantum parts. For instance, the measurement of Pauli observable $Z$ on states $\ket{+}=(\ket{0}+\ket{1})/\sqrt{2}$ and $I/2=(\ket{0}\bra{0}+\ket{1}\bra{1})/2$ both lead to $H(Z)=1$. 

It is natural to consider quantum coherence, which is one of the defining features of quantum mechanics, to quantify the quantum component in uncertainty~\cite{Coles2011,Coles2012,Korzekwa2014}. Along this spirit, rigorous connections between quantum coherence and entropic uncertainty have been established~\cite{Yuan2015,Yuan2019} based on the framework of coherence quantification~\cite{Baumgratz2014}, and the quantum uncertainty relations (QURs) have been theoretically constructed with various coherence measures~\cite{Yuan2017}. On the experimental side, the QURs using relative entropy of coherence have been demonstrated to investigate the trade-off relation~\cite{Lv2018} and connection between entropic uncertainty and coherence uncertainty~\cite{Ding2020}. Still, there are several unexplored matters along the line of experimental investigations. Firstly, although various QURs have been theoretically constructed with relative entropy of coherence, the experimental feasibility and comparison has not been tested. Secondly, the experimental realizations of QURs using other coherence measures beyond relative entropy of coherence is still lacking. Finally, the lower bounds in QURs are generally obtained with quantum state tomography (QST), which becomes a challenge when the dimension of quantum state increases.  

In this paper, we experimentally investigate QURs constructed with three coherence measures, relative entropy of coherence, $l_1$ norm of coherence and coherence of formation, on a family of single-photon states. The lower bound of the QURs are indicated with classical shadow algorithm\cite{huang2020predicting}. We show the tightness of quantum coherence lower bounds depends on the reference bases as well as the purity of quantum state.

This paper is organized as follows. In Section~\ref{sec:QURs} we introduce the basic idea of QUR using quantum coherence measures. In Sections~\ref{sec:CS}, we briefly introduce the CS algorithm to detect purity of a quantum state. In Sections~\ref{sec:exp} and~\ref{sec:expresults}, we present the experimental demonstration and results. Finally, we draw the conclusion in Sections~\ref{sec:conclusion}.

\section{quantum uncertainty relations}
\label{sec:QURs}
A functional $C$ can be regarded as a coherence measure if it satisfies four postulates: non-negativity, monotonicity, strong monotonicity and convexity~\cite{Baumgratz2014}. The different coherence measure plays different roles in quantum information processing. For instance, the relative entropy of coherence plays a crucial role in coherence distillation~\cite{Winter2016}, coherence freezing~\cite{Bromley2015,Yu2016}, and the secrete key rate in quantum key distribution~\cite{Ma2019}. The coherence of formation represents the coherence cost, i.e., the minimum rate of a maximally coherent pure state consumed to prepare the given state under incoherent and strictly incoherent operations~\cite{Winter2016}. The $l_1$-norm of coherence is closely related to quantum multi-slit interference experiments~\cite{Bera2015} and is used to explore the superiority of quantum algorithms~\cite{Hillery2016,Shi2017,Liu2019}. We refer to Ref.~\cite{Streltsov2017} for the review of resource theory of quantum coherence. In the following, we give a brief review of QURs constructed with coherence measures of relative entropy of coherence, $l_1$-norm of coherence, and coherence of formation~\cite{Yuan2017}.

\subsection{QURs using relative entropy of coherence}
The relative entropy of coherence of state $\rho$ is defined as~\cite{Baumgratz2014}
\begin{equation}
    C_{\text{RE}}^{\mathbb{J}}(\rho)=S_{\text{VN}}^{\mathbb{J}}(\rho_d)-S_{\text{VN}}(\rho),
\end{equation}
where $\mathbb{J}=\{\ket{j}\}$ denotes the measurement basis of observable $J$, $S_{\text{VN}}(\rho)=-\tr[\rho\log_2\rho]$ is the von Neumann entropy and $\rho_d$ is the diagonal part of $\rho$ in measurement basis $\mathbb{J}$. Note that $H(J)=S_{\text{VN}}^{\mathbb{J}}(\rho_d)$. The QUR using relative entropy of coherence~\cite{Yuan2017} is
\begin{equation}\label{Eq:QUR of REC yuan}
    C_{\text{RE}}^{\mathbb{A}}(\rho)+C_{\text{RE}}^{\mathbb{B}}(\rho)\geq h\bigg(\frac{\sqrt{2\mathcal{P}-1}(2\sqrt{c}-1)+1}{2}\bigg)-S_\text{VN}(\rho),
\end{equation}
where $h(x)=-x\log_2x-(1-x)\log_2(1-x)$ is the binary entropy and $\mathcal{P}=\tr[\rho^2]$ is the purity of state $\rho$. Similarly, the entropic uncertainty relations proposed by S$\acute{a}$nches-Ruiz\cite{Sanchesruiz1998}, Berta \emph{et al.}\cite{Berta2010} and Korzekwa \emph{et al.}~\cite{Korzekwa2014} can be expressed as
\begin{equation}\label{Eq:QUR of REC Sanchesruiz}
    C_{\text{RE}}^\mathbb{A}(\rho)+C_{\text{RE}}^\mathbb{B}(\rho)\geq h\left(\frac{1+\sqrt{2c-1}}{2}\right)-2S_\text{VN}(\rho),
\end{equation}
\begin{equation}\label{Eq:QUR of REC Berta}
    C_{\text{RE}}^\mathbb{A}(\rho)+C_{\text{RE}}^\mathbb{B}(\rho)\geq-\log_2c-S_\text{VN}(\rho),
\end{equation}
\begin{equation}\label{Eq:QUR of REC Korzekwa}
    C_{\text{RE}}^\mathbb{A}(\rho)+C_{\text{RE}}^\mathbb{B}(\rho)\geq-[1-S_{\text{VN}}(\rho)]\log_2c.
\end{equation}
Consider a qubit state $\rho$ in spectral decomposition $\rho=\lambda\ket{\psi}\bra{\psi}+(1-\lambda)\ket{\psi_\perp}\bra{\psi_\perp}$ with $\lambda$($1-\lambda$) being the eigenvalue associated with eigenvector $\ket{\psi}$($\ket{\psi_\perp}$), we have $S_{\text{VN}}(\rho)=-\lambda\log_2\lambda-(1-\lambda)\log_2(1-\lambda)$ where the purity $\mathcal{P}$ is related to $\lambda$ by $\mathcal{P}=2\lambda^2-2\lambda+1$.

\subsection{QUR of the $l_1$ norm of coherence}
The $l_1$ norm of coherence in a fixed measurement bases $\mathbb{J}$ is defined in the form of 
\begin{equation}
    C_{l_1}^{\mathbb{J}}(\rho)=\sum_{k\neq l}\abs{\bra{j_k}\rho\ket{j_l}},
\end{equation}
and the QUR using $l_1$ norm of coherence is
\begin{equation}\label{Eq:QUR of l1}
    C_{l_1}^{\mathbb{A}}(\rho)+C_{l_1}^{\mathbb{B}}(\rho)\geq 2\sqrt{(2\mathcal{P}-1)c(1-c)}.
\end{equation}

\subsection{QUR using coherence of formation}
The coherence of formation in a fixed measurement bases $\mathbb{J}$ is defined in the form of 
\begin{equation}
    C_f^{\mathbb{J}}(\rho)=\inf_{\{p_i,\ket{\varphi_i}\}}\sum_ip_iC_{\text{RE}}^{\mathbb{J}}(\ket{\varphi_i}\bra{\varphi_i}),
\end{equation}
where the infimum is taken over all state decomposition of $\rho=\sum_{i} p_{i} \ket{\varphi_i}\bra{\varphi_i}$. The QUR using coherence of formation is 
\begin{equation}\label{Eq:QUR of CF}
    C_{f}^\mathbb{A}(\rho)+C_{f}^\mathbb{B}(\rho)\geq h\bigg(\frac{1+\sqrt{1-4(2\mathcal{P}-1)\sqrt{c}(1-\sqrt{c})}}{2}\bigg).
\end{equation}

\section{classical shadow}
\label{sec:CS}
From Section~\ref{sec:QURs}, it is obvious that the purity $\mathcal{P}$ of $\rho$ is the key ingredient in the experimental testing of various QURs. The purity $\mathcal{P}$ can be calculated by reconstructing the density matrix of $\rho$ with QST, which is very costly as the Hilbert space of $\rho$ increases. Another protocol employs two copies of $\rho$ for the detection of $\mathcal{P}$, i.e., $\mathcal P= \tr[\Pi \rho \otimes \rho]$ with $\Pi$ being the local swap operator of two copies of the state~\cite{RevModPhys.81.865,brydges2019probing}.

Very recently, the CS algorithm has been theoretically proposed to efficient quantum state detection~\cite{huang2020predicting}, and has been experimentally realized in the detection of purity of unknown quantum states~\cite{Elben2020,Zhang2021}. In CS algorithm, a randomly selected single-qubit Clifford unitary $U$ is applied on $\rho$, and then the rotated state $U\rho U^{\dagger}$ is measured in the Pauli-$Z$ basis, i.e., $\mathbb{Z}=\{\ket{z_0}=\ket{0},\ket{z_1}=\ket{1}\}$. With the outcome of $\ket{z_i}$, the estimator $\hat{\rho}$ is constructed by $\hat{\rho}= 3U^{\dagger}\ket{z_i}\bra{z_i}U-I$. It is equivalent to measure $J=U^{\dagger}ZU$ ($\mathbb{J}=\{U\ket{0},U\ket{1}\}$) on $\rho$, and the measurement basis $J$ is randomly selected from the Pauli observable basis set $\mathbb{J} \in  \{\mathbb{X},\mathbb{Y},\mathbb{Z}\}$ with an uniform probability $\Kcal(\mathbb{J})=1/3$. The estimator $\hat{\rho}$ can be rewritten as $\hat{\rho}= 3\ket{k}\bra{k}-I$, where $\ket{k}\in \{\ket{x_0}, \ket{x_1}, \ket{y_0}, \ket{y_1}, \ket{z_0}, \ket{z_1}\}$. In particular, $\ket{x_0}=\ket{+}=(\ket{0}+\ket{1})/\sqrt{2}$ and $\ket{x_1}=\ket{-}=(\ket{0}-\ket{1})/\sqrt{2}$ are the eigenvectors of Pauli observable $X$ and $\ket{y_0}=\ket{L}=(\ket{0}+i\ket{1})/\sqrt{2}$ and $\ket{y_1}=\ket{R}=(\ket{0}-i\ket{1})/\sqrt{2}$ are the eigenvectors of Pauli observable $Y$. It is worth noting that the construction of estimator $\hat{\rho}$ only requires one sample. For a set of estimators $\{\hat{\rho_i}\}$ constructed with $N_s$ samples, the purity of state $\rho$ can be estimated by two randomly selected independent $\hat{\rho_i}$ and $\hat{\rho_j}$, i.e. $\hat{\mathcal P}=\sum_{i \neq j}\tr[\hat{\rho_i}\otimes\hat{\rho_j}]/N_s(N_s-1)$.

\section{Experiment realizations}
\label{sec:exp}
\begin{figure}[htb]
\begin{center}
\includegraphics[width=8cm]{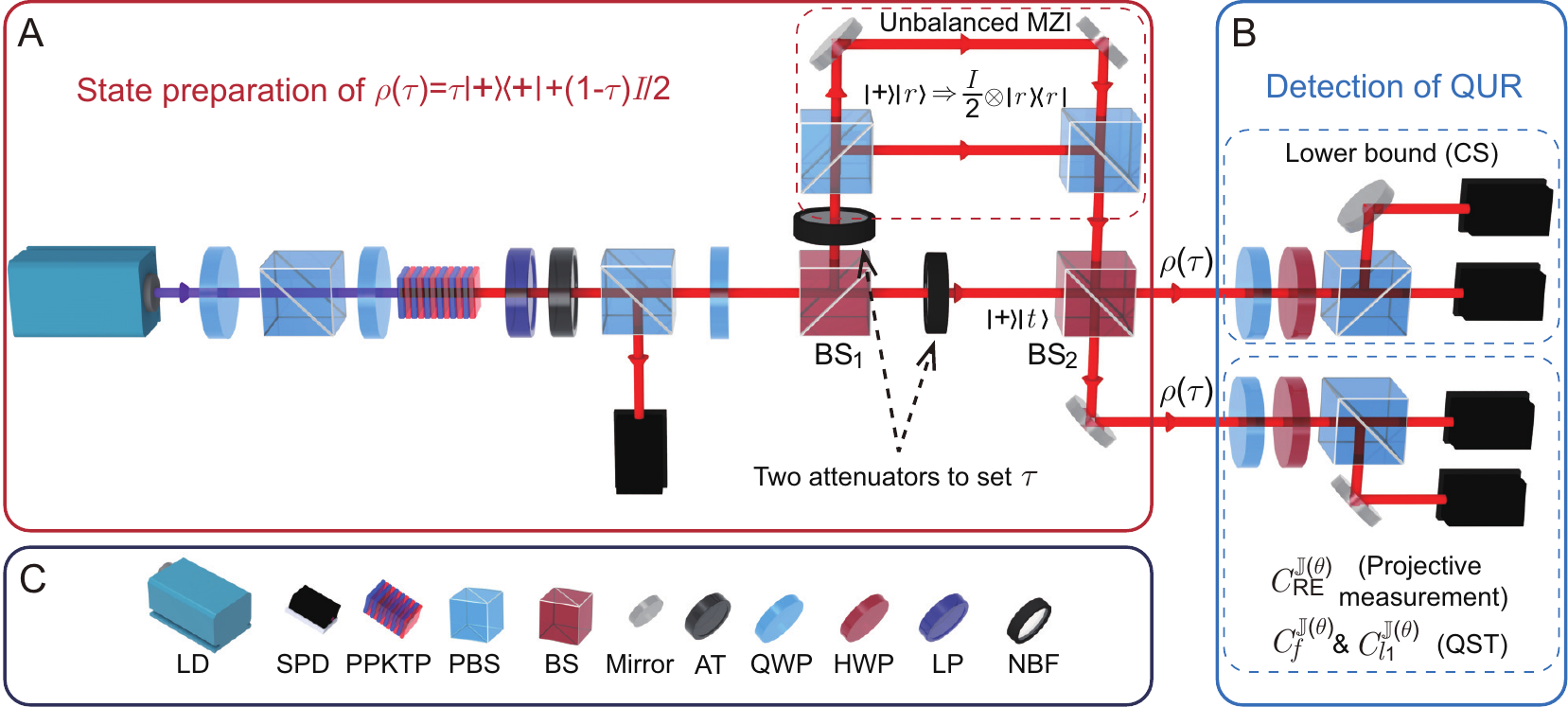}
\end{center}
\caption{Schematic illustration of the experimental setup. \textbf{(A)} The
setup to generate the family of states $\rho(\tau)=\tau\ket{+}\bra{+}+(1-\tau)\frac{I}{2}$.
\textbf{(B)} Experimental setup to implement the measurements with CS algorithm and QST. \textbf{(C)} Symbols used in \textbf{(A)} and \textbf{(B)}. Laser diode (LD); Single-photon detector (SPD); Attenuator (AT); Long-wave pass filter (LP); Narrow-band filter (NBF).}\label{fig:1}
\end{figure}

To test the aforementioned QURs of various coherence measures, we consider the following single-qubit state
\begin{equation}\label{Eq:Expstates}
    \rho(\tau)=\tau\ket{+}\bra{+}+(1-\tau)\frac{I}{2},
\end{equation}
with $0\leq\tau\leq 1$. Note that $\tau=1$ corresponds to the pure state $\ket{+}$ and $\tau=0$ corresponds to the maximally mixed state $I/2$. The experimental setup to generate state in Eq.~\ref{Eq:Expstates} is shown in Figure~\ref{fig:1}A. Two photons are generated on a periodically poled potassium titanyl phosphate (PPKTP) crystal pumped by an ultraviolet CW laser diode. The generated two photons are with orthogonal polarization denoted as $\ket{HV}$, where $\ket{H}$ and $\ket{V}$ denote the horizontal and vertical polarization respectively. Two photons are separated on a polarizing beam splitter (PBS), which transmits $\ket{H}$ and reflects $\ket{V}$. The reflected photon is detected to herald the existence of transmitted photon in state $\ket{H}$, which is then converted to $\ket{+}=(\ket{H}+\ket{V})/\sqrt{2}$ by a half-wave plate (HWP) set at 22.5$^\circ$. We sent the heralded photon into a 50:50 beam splitter (BS$_1$), which transmits (reflects) the single photon with probability of 50\%. The photon in transmitted and reflected mode are denoted as $\ket{t}$ and $\ket{r}$ respectively. Two tunable attenuators are set at modes $\ket{t}$ and $\ket{r}$ to realize the ratio of transmission probability in $\ket{t}$ and $\ket{r}$ of $\frac{\tau}{1-\tau}$. The photon in $\ket{r}$ passes through an unbalanced Mach-Zehnder interferometer (MZI) consisting of two PBS and two mirrors, which acts as a completely dephasing channel in polarization degree of freedom (DOF), i.e., $\ket{+}\bra{+}\to\mathbb{I}/2$. Finally, the two beams are incoherently mixed on BS$_2$ to erase the information of path DOF, which leads to the state $\rho(\tau)$ in both output ports. A step-by-step calculation detailing the evolution of the single-photon state through this setup is given in Eq.~\ref{Eq:Statepreparation} 
\begin{equation}
\begin{split}
\ket{H}&\xrightarrow{\text{HWP}@22.5^\circ}\ket{+}=\frac{1}{\sqrt{2}}(\ket{H}+\ket{V})\\
&\xrightarrow{\text{BS}_1}\ket{+}\otimes\frac{1}{\sqrt{2}}(\ket{t}+\ket{r})\\
&\xrightarrow[\text{at } \ket{t} \text{ and } \ket{r} ]{\text{two attenuators}}\ket{+}\otimes(\sqrt{\tau}\ket{t}+\sqrt{1-\tau}\ket{r})\\
&\xrightarrow[\text{at }\ket{r}]{\text{unbalanced MZI}}\tau\ket{+}\bra{+}\otimes\ket{t}\bra{t}+(1-\tau)I/2\otimes \ket{r}\bra{r}\\
&\xrightarrow[{\text{incoherently combined}}]{\text{BS}_2}\tau\ket{+}\bra{+}+(1-\tau)I/2.\label{Eq:Statepreparation}
\end{split}
\end{equation}

\begin{figure}[htb]
\begin{center}
\includegraphics[width=8cm]{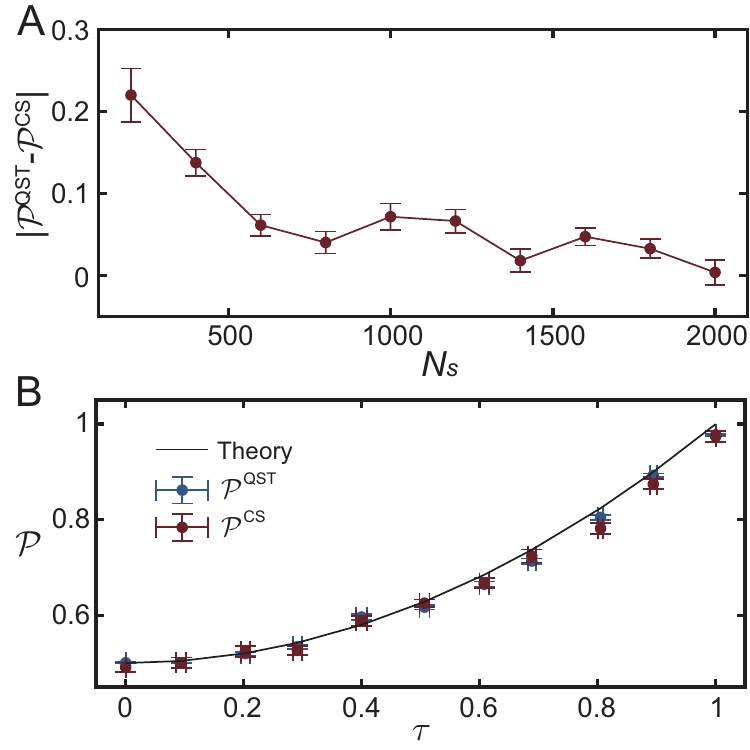}
\end{center}
\caption{\textbf{(A)} The average estimated $\mathcal{P}^{\text{CS}}$ of 11 prepared states with different $N_s$. \textbf{(B)} The results of $\mathcal{P}^{\text{CS}}$ (blue dots) and $\mathcal{P}^{\text{QST}}$ (red dots). The black line is the theoretical prediction of purity of ideal $\rho(\tau)$.}\label{fig:2}
\end{figure} 

\begin{figure*}[ht!b]
\begin{center}
\includegraphics[width=15cm]{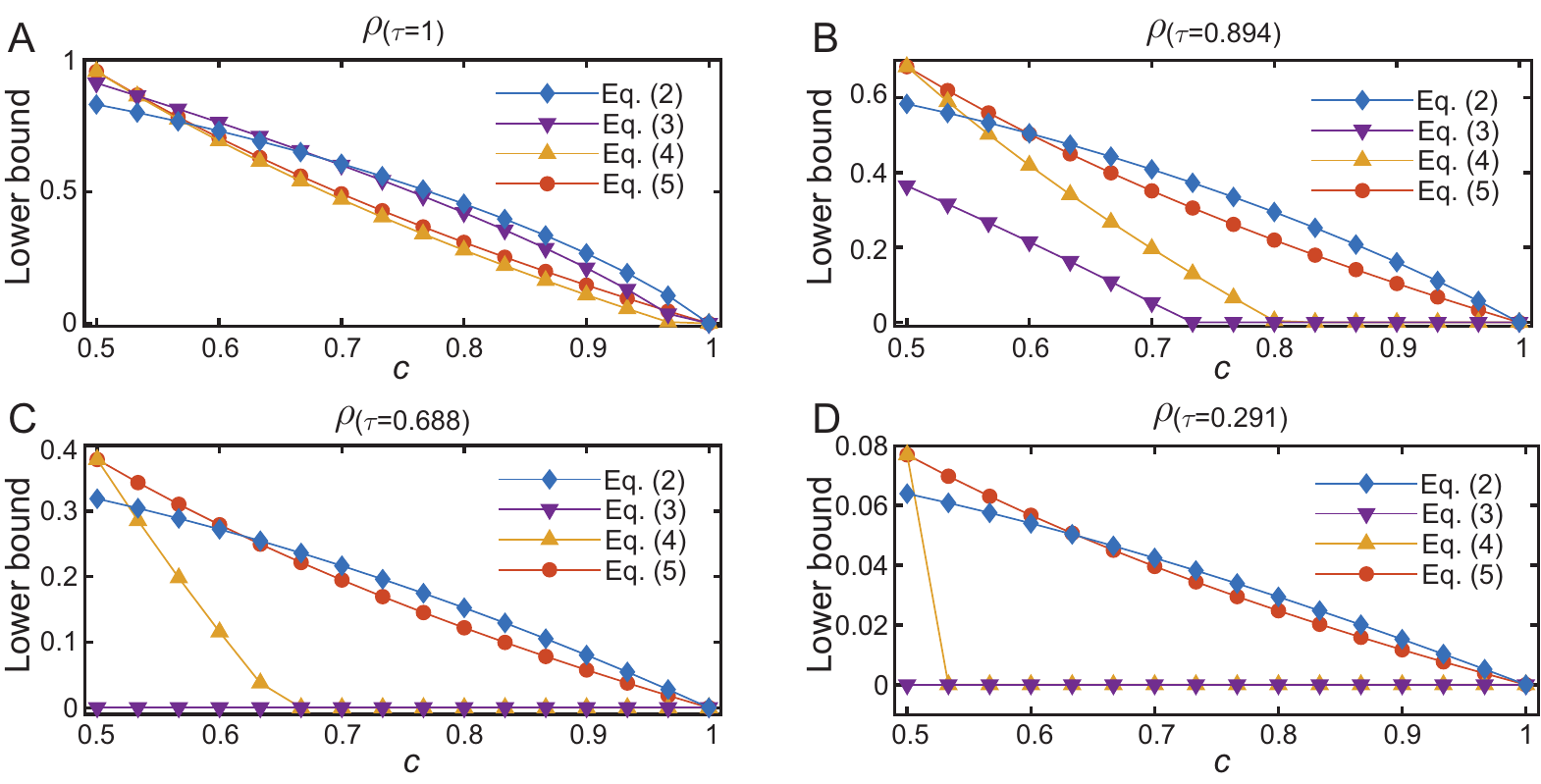}
\end{center}
\caption{The results of estimated lower bounds in Eq.~(2) - Eq.~(5) with different $c$ on state \textbf{(A)} $\rho(\tau=1)$, \textbf{(B)} $\rho(\tau=0.894)$, \textbf{(C)} $\rho(\tau=0.688)$ and \textbf{(D)} $\rho(\tau=0.291)$, respectively.}\label{fig:3}
\end{figure*}

In our experiment, we set the parameter $\tau=0$ to $\tau=1$ with increment of 0.1, and totally generate 11 states. For each generated state, we detect the QURs with setup shown in Figure~\ref{fig:1}B. The lower bound in QURs related to purity $\mathcal{P}^{\text{CS}}$ is measured with CS algorithm. $C^\mathbb{J}_\text{RE}$ is detected with projective measurement on basis $\mathbb{J}$ along with the measured purity. $C^\mathbb{J}_{l_1}$ ($C^\mathbb{J}_f$) is calculated with reconstructed $\rho(\tau)$. All the measurement basis is realized with a HWP, a quarter-wave plate (QWP) and a PBS.

\section{Experimental results}
\label{sec:expresults}
To investigate the accuracy of estimated purity $\mathcal P^{\text{CS}}$ with CS algorithms, we also calculate the purity $\mathcal{P}^{\text{QST}}$ with reconstructed density matrix of $\rho(\tau)$ from QST. The results of $\abs{\mathcal{P}^{\text{QST}}-\mathcal{P}^{\text{CS}}}$ are shown in Figure~\ref{fig:2}A. The more samples used in CS algorithm, the smaller $\abs{\mathcal{P}^{\text{QST}}-\mathcal{P}^{\text{CS}}}$ is. We observe $\abs{\mathcal{P}^{\text{QST}}-\mathcal{P}^{\text{CS}}}<0.1$ when $N_s\geq600$. Especially, $\abs{\mathcal{P}^{\text{QST}}-\mathcal{P}^{\text{CS}}}=0.0036$ when $N_s=2000$. In Figure~\ref{fig:2}B, we shown the results of $\mathcal P^{\text{CS}}$ with $N_s=2000$ and $\mathcal P^{\text{QST}}$  on 11 prepared $\rho(\tau)$, in which the experimental results of $\mathcal P^{\text{CS}}$ and $\mathcal P^{\text{QST}}$ have good agreements with the theoretical predictions. In the following, all the results with CS algorithm are obtained with 2000 samples.

\begin{figure*}[hbt!]
\begin{center}
\includegraphics[width=15cm]{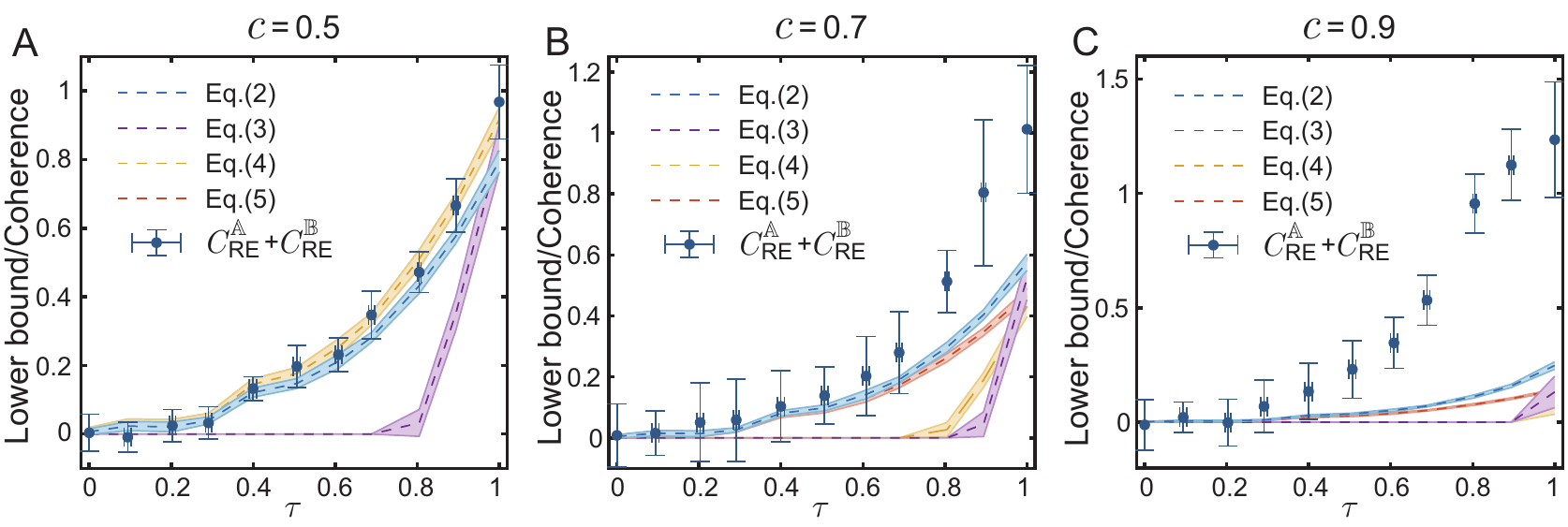}
\end{center}
\caption{The results of QURs in Eq.~(2) - Eq.~(5) on 11 prepared states with \textbf{(A)} $c=0.5$, \textbf{(B)} $c=0.7$ and \textbf{(C)} $c=0.9$. the dashed lines are the measured lower bounds and the shadow area represents the statistical error by repeating CS measurement for 20 times. }\label{fig:4}
\end{figure*}

\begin{figure*}[hbt!]
\begin{center}
\includegraphics[width=15cm]{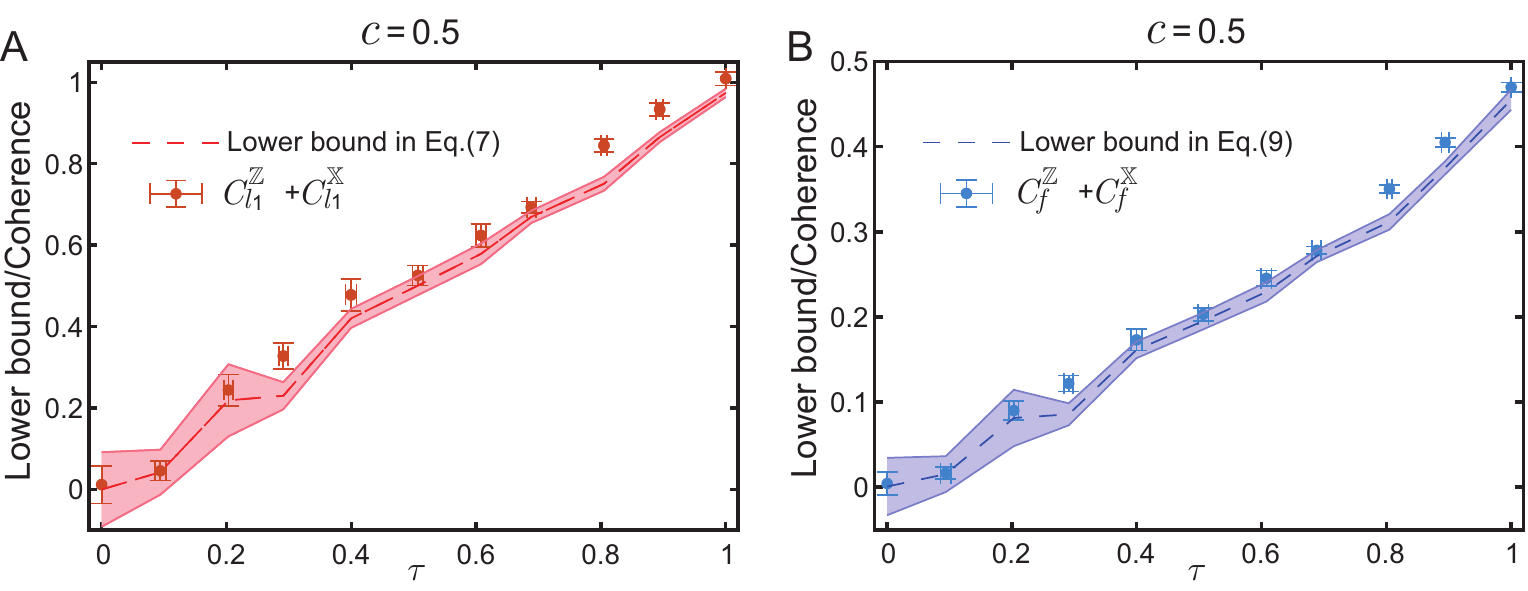}
\end{center}
\caption{The results of \textbf{(A)} QUR with $l_1$ norm of coherence and \textbf{(B)} QUR with coherence of formation with $c=0.5$.}\label{fig:5}
\end{figure*} 

We first focus on the QURs using relative entropy of coherence, i.e., Eq.~\ref{Eq:QUR of REC yuan}-Eq.~\ref{Eq:QUR of REC Korzekwa}. We calculate the lower bounds in Eq.~\ref{Eq:QUR of REC yuan}-Eq.~\ref{Eq:QUR of REC Korzekwa} with the estimated $\mathcal P^{\text{CS}}$ on $\rho(\tau=1)$, $\rho(\tau=0.894)$, $\rho(\tau=0.688)$ and $\rho(\tau=0.291)$, respectively. As shown in Figure~\ref{fig:3}A, we observe the lower bounds in Eq.~\ref{Eq:QUR of REC Berta} and Eq.~\ref{Eq:QUR of REC Korzekwa} 
have the same value when $\mathbb{A}$ and $\mathbb{B}$ are mutually unbiased ($c=0.5$), and outperform others. When $c$ becomes larger, lower bounds in Eq.~\ref{Eq:QUR of REC yuan} and Eq.~\ref{Eq:QUR of REC Sanchesruiz} are stricter than that in ~\ref{Eq:QUR of REC Berta} and Eq.~\ref{Eq:QUR of REC Korzekwa}. However, the situation is quite different when the purity becomes smaller.  As shown in  Figure~\ref{fig:3}B-Figure~\ref{fig:3}D, the values of lower bounds in Eq.~\ref{Eq:QUR of REC Sanchesruiz} and Eq.~\ref{Eq:QUR of REC Berta} are negative (we denote them as 0) when $c$  is larger than certain values, which means the lower bounds are loosen as $ C_{\text{RE}}^\mathbb{A}(\rho)+C_{\text{RE}}^\mathbb{B}(\rho)>0$ for all $\rho$.

To investigate the tightness of various lower bounds, we measure $C_{\text{RE}}^\mathbb{A}(\rho)+C_{\text{RE}}^\mathbb{B}(\rho)$ in different reference bases. We select observables $A$ and $B$ from set $J(\theta)=\cos\theta Z+\sin\theta X$. Specifically, we fix $A=J(0^\circ)$ and choose $B=J(90^\circ), J(66.42^\circ)$ and $J(36.86^\circ)$, which corresponds to $c=0.5, 0.7$ and $0.9$. For each observable $J(\theta)$, we preform the projective measurement on basis $\mathbb{J}(\theta)$, and calculate the Shannon entropy of measurement outcomes $H(J(\theta))$. Thus, we obtain $C_{\text{RE}}^{\mathbb{J}(\theta)}(\rho(\tau))=H(J(\theta))-S_{\text{VN}}(\rho(\tau))$, where $S_{\text{VN}}(\rho(\tau))$ can be calculated from $\mathcal{P}^{\text{CS}}$. The results of QURs using relative entropy of coherence are shown in Figure~\ref{fig:4}. As shown in Figure~\ref{fig:4}A, the lower bounds in Eq.~\ref{Eq:QUR of REC Berta} and Eq.~\ref{Eq:QUR of REC Korzekwa} have the same values as $ C_{\text{RE}}^\mathbb{A}(\rho)+C_{\text{RE}}^\mathbb{B}(\rho)$ is lower bounded by $1-S_\text{VN}(\rho)$ when $c=0.5$ according to the definitions in Eq.~\ref{Eq:QUR of REC Berta} and Eq.~\ref{Eq:QUR of REC Korzekwa}. When $c$ is larger, the lower bound in Eq.~\ref{Eq:QUR of REC yuan} is stricter than others as reflected in Figure~\ref{fig:4}B and Figure~\ref{fig:4}C.

Next, we investigate the QURs using coherence of formation and $l_1$-norm of coherence as described in Eq.~\ref{Eq:QUR of CF} and Eq.~\ref{Eq:QUR of l1}. We choose observables $A=J(0^\circ)=Z$ and $B=J(90^\circ)=X$ in the coherence measure, which corresponds to $c=0.5$. The $C_{l_1}^\mathbb{Z}(\rho)$ and $C_{l_1}^\mathbb{X}(\rho)$ are calculated according to Eq.~\ref{Eq:QUR of l1} with the reconstructed density matrix of $\rho(\tau)$. Thus, $C_f^\mathbb{Z}(\rho)$ and $C_f^\mathbb{X}(\rho)$ can be calculated with $C_{l_1}^\mathbb{Z}(\rho)$ and $C_{l_1}^\mathbb{X}(\rho)$ as $C_f(\rho)=h\left(\frac{1+\sqrt{1-C_{l_1}(\rho)}}{2}\right)$\cite{Yuan2017}. The results of QURs using $l_1$ norm of coherence and coherence of formation are shown in Figure~\ref{fig:5}A and Figure~\ref{fig:5}B respectively, in which the measured coherence are well bounded by the measured lower bounds.

\section{Conclusion}
\label{sec:conclusion}
In this paper, we experimentally investigate quantum uncertainty relations using various coherence measures. The lower bounds in quantum uncertainty relations are detected with classical shadow algorithm, in which the measurement cost is quite small and independent of the dimension of quantum states. For the quantum uncertainty relation using relative entropy of coherence, we show that the tightness of lower bounds is highly related to the reference basis as well as purity of quantum state. Moreover, we test the quantum uncertainty relation using $l_1$ norm of coherence and coherence of formation. 

Our results could benefit the choice of quantum uncertainty relations using quantum coherence in practice, especially when considering the inevitable experimental imperfections. More importantly, our method can be generalized to multipartite states while keeps its efficiency. The multiparty coherence could be efficiently estimated using stablizer theory~\cite{Ding2021PRR,Ding2021Entropy}, and the classical shadow algorithm to detect purity of multipartite state is efficient as well~\cite{Zhang2021}.

\bibliography{CUR.bib}

\end{document}